\documentclass[aps,prl,twocolumn,superscriptaddress]{revtex4-1}
\usepackage{graphicx}
\usepackage{amsmath}
\begin{document}


\title{Asymptotic behavior of the solution of the space-dependent variable-order fractional diffusion equation: ultra-slow anomalous aggregation}

\author{Sergei Fedotov}
\author{Daniel Han}
\affiliation{School of Mathematics, University of Manchester, M13 9PL}

\date{\today}

\begin{abstract}
We find the asymptotic representation of the solution of the variable-order fractional diffusion equation, which remains unsolved since it was proposed in [Checkin \textit{et. al.}, \textit{J. Phys. A}, 2005]. We identify a new advection term that causes ultra-slow spatial aggregation of subdiffusive particles due to dominance over the standard advection and diffusion terms, in the long-time limit. This uncovers the anomalous mechanism by which non-uniform distributions can occur. We perform Monte Carlo simulations of the underlying anomalous random walk and find good agreement with the asymptotic solution.
\end{abstract}


\maketitle

Anomalous diffusion has attracted immense interest in the past due to many physical, chemical and biological processes characterized by the mean square displacement (MSD) involving the fractional exponent $\mu$: $\langle x^2(t)\rangle\propto t^{\mu}$ \cite{metzler1999anomalous,metzler2000random,klages2008anomalous,mendez2010reaction,klafter2011first,henry2010fractional}. Anomalous diffusion is observed also in many other areas, for instance, in finance and economics \cite{scalas2006application}.
An influential paper by Metzler and Klafter \cite{metzler2000random} reviews anomalous diffusion in the scope of a constant  exponent $\mu$.
However, anomalous transport in realistic inhomogeneous and complex environments \cite{lanoiselee2018diffusion}, such as lipid granules \cite{jeon2011vivo}, porous media \cite{edery2015anomalous} and entangled polymer liquids \cite{cai2015hopping}, requires a multi-fractional approach involving the space-dependent variable-order fractional exponent \cite{chechkin2005fractional,sun2009variable,korabel2010paradoxes,fedotov2012subdiffusive,straka2018variable,berry2014spatial,kian2018time}.  Important examples  of anomalous transport involving multi-fractional exponents are lateral diffusion of proteins on crowded lipid membranes \cite{jeon2016protein}, intracellular subdiffusion of proteins \cite{weiss2004anomalous}, mRNA \cite{golding2006physical} and organelles \cite{korabel2018non} due in part to inhomogeneous crowding \cite{ghosh2016anomalous} and weak interactions between components in the cell \cite{ba2018whole}. Recent observations show that lysosomes, which are key organelles for cellular metabolism, predominantly move subdiffusively and maintain a non-uniform spatial distribution in the cell \cite{ba2018whole}. The majority of these organelles are concentrated in the perinuclear area. A fundamental unresolved question is how lysosomes are self-organized spatially to coordinate their roles \cite{ba2018whole}. In this Letter, we propose a new anomalous mechanism by which non-uniform distribution of subdiffusing organelles can occur.

A generic model for anomalous diffusion in inhomogeneous media is the space-dependent variable-order fractional diffusion equation \cite{chechkin2005fractional,sun2009variable,korabel2010paradoxes,fedotov2012subdiffusive,straka2018variable}
\begin{equation}
\frac{\partial p(x,t)}{\partial t} = \frac{\partial^2}{\partial x^2}\left[ D_{\mu(x)} \mathcal{D}_t^{1-\mu(x)} p(x,t) \right] ,
\label{sdvofde}
\end{equation}
where $p(x,t)$ is the probability density function (PDF) of a particle at position $x$ and time $t$. This function can be also interpreted as the mean number density of subdiffusive particles.
In Eq. (\ref{sdvofde}), $D_{\mu(x)} = a^2/2\tau_0^{\mu(x)}$ is the fractional diffusion coefficient with the microscopic time scale $\tau_0$, length scale $a$, and space-dependent fractional exponent $\mu(x) \in (0,1)$. The Riemann-Liouville derivative 
\begin{equation}
	\mathcal{D}_t^{1-\mu(x)}p(x,t) = \frac{1}{\Gamma(\mu(x))}\frac{\partial}{\partial t}\int_{0}^{t} \frac{p(x,t')}{(t-t')^{1-\mu(x)}}dt'
\end{equation} 
also involves spatial dependence. Equation (\ref{sdvofde}) was first derived by Chechkin, Gorenflo and Sokolov \cite{chechkin2005fractional}, and since then, many attempts have been made to find a solution through composite regions with constant anomalous exponents and numerically \cite{chechkin2005fractional,chen2010numerical,korabel2010paradoxes}. However, Eq. (\ref{sdvofde}) remains unsolved for the general case of a space-dependent anomalous exponent $\mu(x)$. 

\begin{figure}[]
	\centering
	\includegraphics[width=\linewidth]{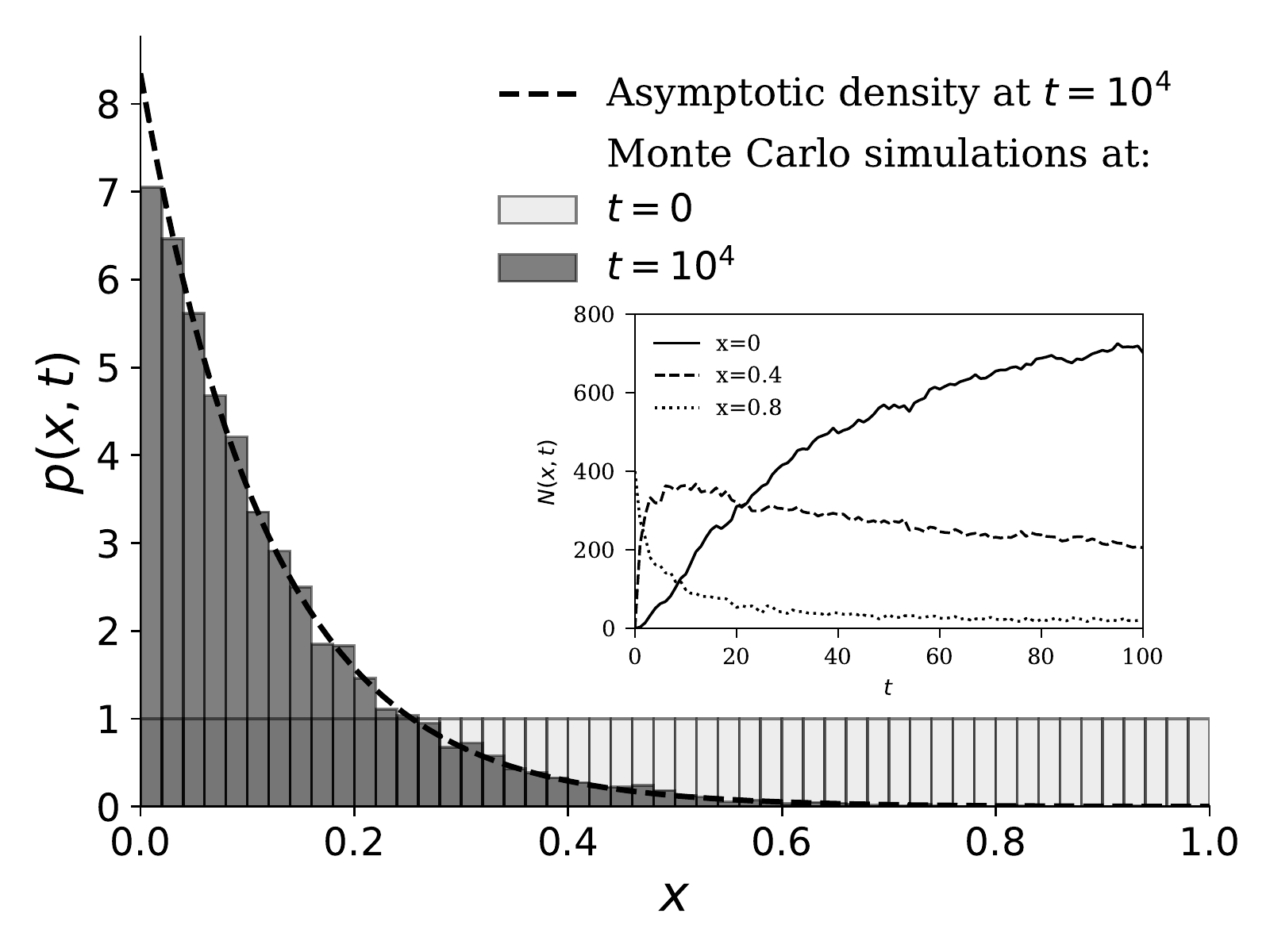}
	\caption{Asymptotic density (\ref{longtsolfde}) (dashed line) and normalized histograms corresponding to simulation of $N = 10^4$ particles jumping between $k = 50$ bins in the domain $0\leq x\leq 1$ with fractional exponent $\mu(x) = 0.4 + 0.5x$ and $\tau_0 = 10^{-3}$. In this simulation, $r_i = 0.5$. The legend shows simulation times, $t$, at which snapshots of the distribution of particles was produced. \textit{Inset:} Time series of number of particles $N(x,t)$ at specific $x$ positions for the same simulation in the main figure. The plot at $x=0$ clearly shows logarithmic growth as predicted by solution (\ref{longtsolfde}).}
	\label{fig:fig1}
\end{figure}

In this Letter, we find the asymptotic representation of the solution of the space-dependent variable-order fractional diffusion equation (\ref{sdvofde}) for a monotonically increasing fractional exponent. In the long-time limit, we obtain the normalized density
\begin{equation}
p(x,t) \sim \frac{\mu'_0\left(\frac{t}{\tau_0}\right)^{-\Delta\mu( x)}}{\Gamma\left(1-\Delta\mu( x)\right)}\left[ \ln\left(\frac{t}{\tau_0}\right) - \psi_0\left(1-\Delta\mu( x)\right) \right].
\label{longtsolfde}
\end{equation}
 This asymptotic density is in the domain $0<x<L$ with reflective boundary conditions, subject to $\mu'_0 = \frac{d\mu}{dx}(0)\neq 0$, where $\Delta\mu(x) = \mu(x)-\mu(0)$ and $\psi_0(x) = \Gamma'(x)/\Gamma(x)$ is the digamma function. For linearly increasing $\mu(x)$, we have $\Delta\mu(x) = \mu'_0 x$ and $\mu'_0 = \left[\mu(L)-\mu(0)\right]/L$. The unsteady non-uniform distribution (\ref{longtsolfde}) for $t=10^4$ is illustrated by the dashed line in Fig. \ref{fig:fig1}.

The unusual feature of this unsteady representation is that it describes ultra-slow formation of a non-uniform distribution of subdiffusive particles (spatial aggregation). It follows from (\ref{longtsolfde}) that $p(x,t)$ at $x=0$ is $\mu_0' \left[ \ln (t/\tau_0) + \gamma \right]$ where $\gamma$ is the Euler-Mascheroni constant, which results in ultra-slow aggregation at the minimum value of $\mu(x)$ as seen in Fig.\ref{fig:fig1}. In fact, $p(x,t)$ tends to delta-function $\delta (0)$ \cite{fedotov2012subdiffusive} but it takes an extremely long time due to the logarithmic growth. 

This asymptotic behavior of the solution (\ref{longtsolfde}) can be explained by the anomalous continuous time random walk (CTRW) where the fractional exponent $\mu(x)$ is a measure of the trapping strength. This is because the waiting time density of underlying random walkers is given by $\psi(\tau,x) \sim 1/\tau^{1+\mu(x)}$ and so the smaller the value of $\mu(x)$, the more likely that the random walker at point $x$ waits longer until the next jump. Therefore, it is expected that eventually the random walkers become trapped in the position with the lowest $\mu(x)$ \cite{korabel2010paradoxes,fedotov2012subdiffusive}. The ultra-slow relaxation is due to the fractional exponent $\mu(x)$ changing in a continuous fashion. 

This behaviour is fundamentally different from the standard formation of non-uniform distributions described by steady-state solutions for advection-diffusion equations \cite{schnitzer1993theory,othmer1997aggregation}. In particular, the Markovian analog of Eq. (\ref{sdvofde}),   $ \partial p / \partial t = \partial^2/\partial x^2\left[D(x) p\right]$ (see (17) in \cite{othmer1997aggregation}), under reflecting boundary conditions, has a stationary solution of $p(x) = A/D(x),$ where $A$ is the normalization constant. This non-uniform steady-state solution occurs as a result of balance between the drift (advection) term $(\partial D(x) /\partial x) p$ and diffusion $D(x)\partial p/\partial x$.

However, for Eq. (\ref{sdvofde}), the mechanism for formation of a non-uniform distribution is very different. To elucidate the origin of this anomalous mechanism, we rewrite Eq. (\ref{sdvofde}) in the form $\partial p/\partial t = -\partial J/\partial x$ with the flux 
\begin{equation*}
	J(x,t) = -\frac{\partial}{\partial x} \left[ \frac{a^2}{2\tau_0^{\mu(x)}} \frac{1}{\Gamma(\mu(x))}\frac{\partial}{\partial t}\int_{0}^{t} \frac{p(x,t')}{(t-t')^{1-\mu(x)}}dt' \right]
	.
\end{equation*}
By differentiating w.r.t $x$, one can obtain the flux $J(x,t)$ as a combination of spatially varying advection and diffusion terms. Explicitly, 
\begin{equation}
\begin{aligned}
J(x,t) = {} &\frac{d\mu}{dx}\ln(\tau_0) D_{\mu(x)}\mathcal{D}_t^{1-\mu(x)}p(x,t)\\
&+ \psi_0(\mu(x))\frac{d\mu}{dx}D_{\mu(x)}\mathcal{D}_t^{1-\mu(x)}p(x,t)\\
&-\frac{d\mu}{dx}D_{\mu(x)}\frac{1}{\Gamma(\mu)}\frac{\partial}{\partial t}\int_{0}^{t} \frac{\ln\left(t-t'\right)}{(t-t')^{1-\mu}}p(x,t')dt'\\
& -D_{\mu(x)}\mathcal{D}_t^{1-\mu(x)}\frac{\partial p(x,t)}{\partial x}.
\label{longJ}
\end{aligned}
\end{equation}
Then combining the logarithms in the first and third term and defining a fractional operator $\mathcal{U}_t^{1-\mu} p$, we can write more neatly
\begin{multline}
J(x,t) = -D_{\mu(x)}\mathcal{D}_t^{1-\mu(x)} \frac{\partial p}{\partial x}\\
\raggedright
-D_{\mu(x)}\frac{d\mu}{dx} \left[  \mathcal{U}_t^{1-\mu(x)}p -
 \psi_0(\mu(x))\mathcal{D}_t^{1-\mu(x)}p\right].
\label{fluxfractionaladvection}
\end{multline}
Here $\mathcal{D}_t^{1-\mu(x)}$ is the same operator as in (\ref{sdvofde}), $\psi_0(\cdot)$ is the digamma function, and $\mathcal{U}_t^{1-\mu(x)} p$ is a fractional operator defined as
\begin{equation}
\mathcal{U}_t^{1-\mu(x)} p = \frac{1}{\Gamma(\mu)}\frac{\partial}{\partial t}\int_{0}^{t} \frac{\ln\left((t-t')/\tau_0\right)}{(t-t')^{1-\mu(x)}}p(x,t')dt'.
\label{modifiedRLderivative}
\end{equation}
This operator occurs as a result of space-dependent fractional exponent $\mu(x)$. One can see that it is a modification of the Riemann-Liouville derivative with a logarithmic factor in the memory kernel $\ln\left((t-t')/\tau_0\right)$. The Laplace transform of $\mathcal{U}_t^{1-\mu} p$ can be found by using the convolution theorem and the formula $\mathcal{L}\left\lbrace \ln(t)/t^{1-\mu} \right\rbrace = \Gamma(\mu)\left[\psi_0(\mu)-\ln(s)\right]/s^\mu$ (see \cite{jeffrey2007table}, pp.573):
\begin{equation}
\mathcal{L}  \left\lbrace \mathcal{U}^{1-\mu}_t p \right\rbrace = \left[ \psi_0(\mu)-\ln(\tau_0 s) \right] s^{1-\mu}\hat{p}(x,s).
\label{modifiedRLlaplace}
\end{equation}


We should note that the flux $J(x,t)$ in (\ref{fluxfractionaladvection}) results from a choice of fractional diffusion equation (\ref{sdvofde}), which is not unique. The form of the coarse-grained fractional equations depend on the microscopic picture of the underlying random walk (see a similar discussion for the Markovian case in Ref. \cite{schnitzer1993theory,othmer1997aggregation}). To illustrate how the fractional equation changes due to underlying microscopic mechanisms, consider symmetric anomalous random walks on a lattice, with spacings of size $a$. The master equation is $\partial p /\partial t = -i(x,t) + i(x-a,t)/2 + i(x+a,t)/2$, where the escape rates $i(x,t)$ from a trap at position $x$ is defined locally such that 
\begin{equation}
	i(x,t) = \frac{1}{\tau_0^{\mu(x)}} \mathcal{D}^{1-\mu(x)}_t p(x,t)
\end{equation}
(see Eq. 30 in Ref. \cite{fedotov2012subdiffusive}). In the limit $a\rightarrow0$ and $\tau_0\rightarrow0$ such that $D_{\mu(x)} = a^2/2\tau_{0}^{\mu(x)}$ is finite, we obtain fractional diffusion equation (\ref{sdvofde}). However, if we introduce escape rates $i^{\pm}(x,t)$ on the right ($+$) and the left ($-$) depending on the barriers at $x\pm a/2$, then the corresponding master equation is 
\begin{equation}
	\frac{\partial p(x,t)}{\partial t} = -i^-(x,t)-i^+(x,t)+i^-(x+a,t)+i^+(x-a,t),
	\label{barrierme}
\end{equation}
where
\begin{equation}
i^{\pm}(x,t) = \frac{1}{2 \tau_0^{\mu(x\pm a/2)}} \mathcal{D}^{1-\mu(x\pm a/2)}_t p(x,t).
\end{equation}
In the limit $a\rightarrow0$ and $\tau_0\rightarrow0$, the master equation (\ref{barrierme}) becomes
\begin{equation}
	\frac{\partial p(x,t)}{\partial t} = \frac{\partial}{\partial x} \left[	D_{\mu(x)} \mathcal{D}_t^{1-\mu(x)} \frac{\partial p}{\partial x} 	\right]
\end{equation}
and the flux is $J(x,t)=-D_{\mu(x)} \mathcal{D}_t^{1-\mu(x)} \partial p(x,t)/\partial x$. Clearly, there is no advection for this case and, instead of our solution (\ref{longtsolfde}), $p(x,t)$ tends to a uniform distribution $1/L$ as $t\rightarrow\infty$. So the conclusion is that for space-dependent anomalous exponent, we cannot rely on phenomenological arguments and need microscopic random walk models to determine the coarse-grained fractional governing equations. A similar situation occurs when the Fokker-Planck equation is derived from the Langevin equation with multiplicative noise \cite{sokolov2010ito}. It follows from solution (\ref{longtsolfde}) that fractional equation (\ref{sdvofde}) describes anomalous transport in non-equilibrium systems, for which long-time behavior does not correspond to Boltzmann equilibrium.

For positive values of $\frac{d\mu}{dx}$, the advection term in (\ref{fluxfractionaladvection}) encapsulates the drift of particles towards the region of lowest $\mu(x)$. The surprising property of this advection term is that it is always dominant, regardless of the value of the gradient $\frac{\partial p}{\partial x}$, in the long-time limit and can never be balanced by diffusion. In other words, there exists no steady-state solution for the diffusion equation with flux (\ref{fluxfractionaladvection}) as $t$ tends to infinity.
Let us demonstrate the dominance of the anomalous advection term by taking the Laplace transform of $\partial p/\partial t = -\partial J/\partial x$ which leads to
\begin{equation}
	s\hat{p}(x,s) - p(x,0) =  -\frac{\partial \hat{J}(x,s)}{\partial x},
	\label{laplacefde}
\end{equation}
where the Laplace transform of the flux is
\begin{equation}
\hat{J}(x,s)=-\frac{a^2s}{2\left(\tau_0s\right)^{\mu(x)}}\left[ \frac{\partial \hat{p}}{\partial x} - \ln(\tau_0s)\frac{d\mu}{dx}\hat{p} \right].
\label{laplaceflux}
\end{equation}
In the limit $s\rightarrow0$, the left hand side of Eq. (\ref{laplacefde}) becomes negligible compared to the right hand side. Therefore, we equate $\hat{J}(x,s)$ to zero and obtain
\begin{equation}
\frac{\partial \hat{p}(x,s)}{\partial x} = \ln(\tau_0s)\frac{d\mu}{dx}\hat{p}(x,s).
\end{equation}
It is clear that as $s\rightarrow 0$, the logarithmic factor $\ln(\tau_0 s)$ on the right hand side tends to $-\infty$, which explains the dominance of the advection in the long-time limit.
The solution to this equation with the normalization condition is
\begin{equation}
	s\hat{p}(x,s) = \frac{\left(\tau_0s\right)^{\mu(x)}}{\int_{0}^{L}\left(\tau_0s\right)^{\mu(x)}dx} .
	\label{laplacesolution}
\end{equation}
Since $\mu(x)$ is an increasing function and it has a minimum at $x=0$, as $s\rightarrow0$, the peak of $(\tau_0s)^{\mu(x)} = \exp\left[\mu(x) \ln(\tau_0s)\right]$ is concentrated in the neighborhood of $x=0$. So we can use the Laplace method to obtain $\int_{0}^{L}\left(\tau_0s\right)^{\mu(x)}dx \sim -(\tau_0 s)^{\mu(0)}/\left[\mu'_0\ln(\tau_0s)\right]$.
Therefore,
 \begin{equation}
 	s\hat{p}(x,s) \sim -(\tau_0s)^{\Delta\mu(x)}\mu'_0 \ln(\tau_0s).
 \end{equation}
Taking the inverse Laplace transform, we obtain the asymptotic density (\ref{longtsolfde}). This asymptotic form is a result of an anomalous aggregation mechanism with a dominant advection term, which has no analogue in classical advection-diffusion equations.

In fact, the anomalous advection term in Eq. (\ref{fluxfractionaladvection}) is so dominant that it overpowers the standard drift such that instead of an equilibrium Boltzmann distribution, Eq. (\ref{longtsolfde}) becomes the asymptotic solution of the space-dependent variable-order fractional Fokker-Planck equation \cite{fedotov2012subdiffusive,straka2018variable}:
in the long-time limit. So Eq. (\ref{longtsolfde}) remains a valid asymptotic representation of the solution for the general space-dependent variable-order fractional Fokker-Planck equation \cite{straka2018variable}.
\begin{equation}
	\frac{\partial p}{\partial t} = -\frac{\partial}{\partial x}\left[ v(x)D_{\mu(x)}\mathcal{D}_t^{1-\mu(x)}p-
	 \frac{\partial}{\partial x}D_{\mu(x)} \mathcal{D}_t^{1-\mu(x)} p \right],
	 \label{voffpe}
\end{equation}
where the drift function, $v(x)= \frac{2(r(x)-l(x))}{a}$, can be found from the non-symmetrical random walk on a lattice with the space distance $a$; $r(x)$ is the probability of particles at position $x$ moving right; and $l(x) = 1-r(x)$ is the probability moving left.

\begin{figure}[h!]
	\centering
	\includegraphics[width=\linewidth]{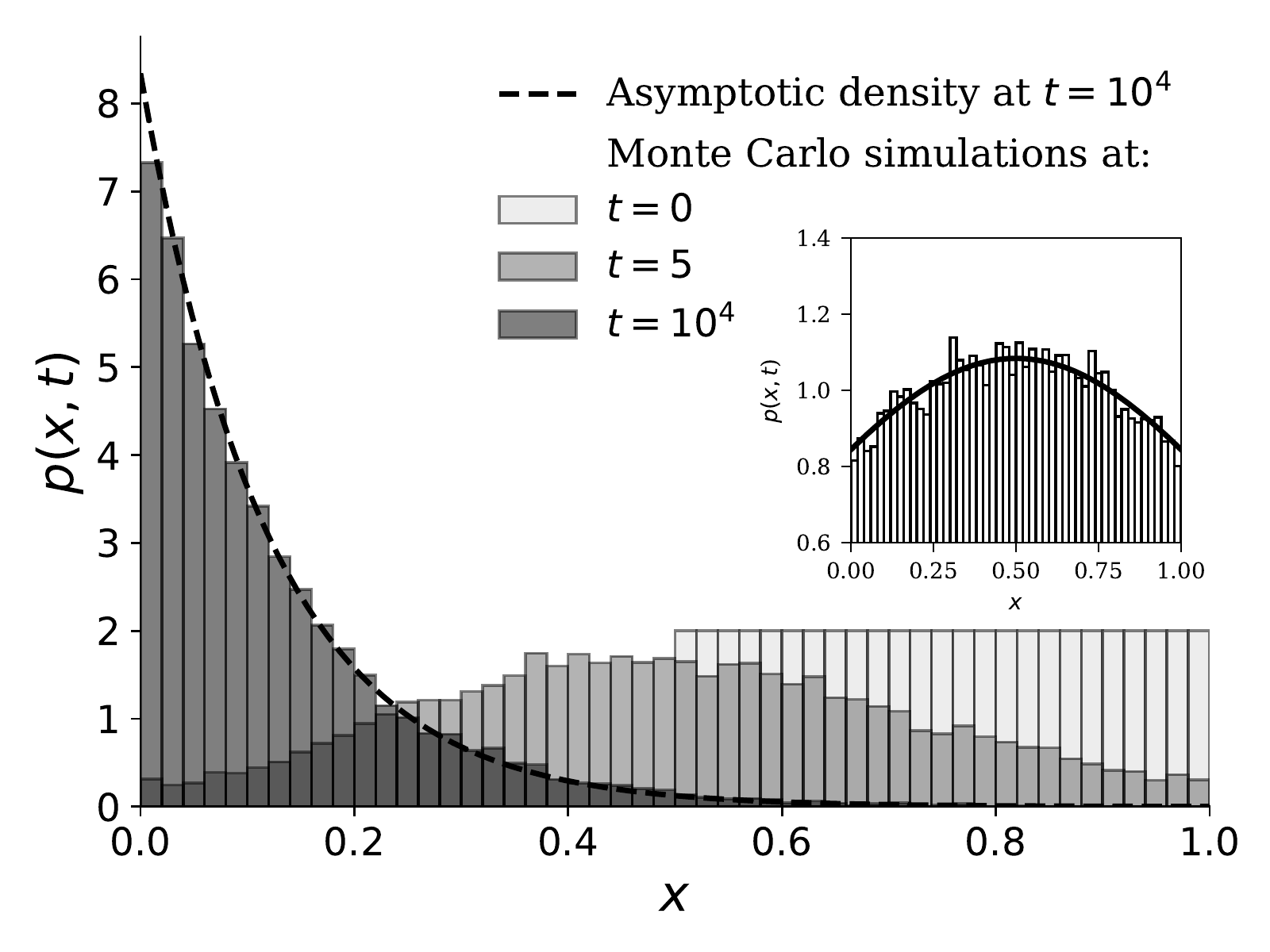}
	\caption{\textit{Main:} Asymptotic density (\ref{longtsolfde}) (dashed line) and normalized histograms corresponding to simulation of $N= 10^4$ particles jumping between $k = 50$ bins in the domain $0\leq x\leq 1$ with fractional exponent $\mu(x) = 0.4 + 0.5x$ and $\tau_0 = 10^{-3}$. In this simulation, $r_i = \frac{1}{2}+\frac{0.5}{k}\left( 0.5 - \frac{i}{k} \right)$. The legend shows simulation times, $t$, at which snapshots of the distribution of particles was produced. \textit{Inset:} The steady solution of the fractional Fokker-Planck equation for constant $\mu$ and $v(x) = 1-2x$: $p(x) = Ce^{-(x^2-x)}$ (solid line) and normalized histogram of particles at $t = 10^4$ with the same parameters and initial conditions as the main histograms except with $N =5\times 10^4$ and mean fractional exponent $\bar{\mu} = \int_{0}^{1} \mu(x)dx = 0.65$.}
	\label{fig:fig2}
\end{figure}

To show the dominance over the standard drift, we take the Laplace transform of Eq. (\ref{voffpe}). The equation will be the same as (\ref{laplacefde}) but with a modified flux
\begin{equation}
	\hat{J}(x,s) = -\frac{a^2s}{2\left(\tau_0s\right)^{\mu(x)}}\left[ \frac{\partial \hat{p}}{\partial x} +v(x)\hat{p} - \ln(\tau_0s)\frac{d\mu}{dx}\hat{p} \right].
	\label{modifiedfluxffpe}
\end{equation}
Just as before, in the long-time limit as $s\rightarrow0$ and $\ln(\tau_0s)\rightarrow-\infty$, the advection term $v(x) \hat{p}$ in Eq. (\ref{modifiedfluxffpe}) is negligibly small compared to the advection $-\ln(\tau_0s)\frac{d\mu}{dx} \hat{p}$ generated by the non-uniform nature of the anomalous exponent $\mu(x)$. Therefore, Eq. (\ref{longtsolfde}) is also the long-time asymptotic representation of the solution to Eq. (\ref{voffpe}). This is confirmed by Monte Carlo simulation shown in Fig.\ref{fig:fig2}.

Another measure which again demonstrates the ultra-slow formation of a non-uniform distribution is the mean position $\bar{x}(t)=\int_{0}^{L} xp(x,t) dx$. Using Eq. (\ref{longtsolfde}), we find
$\bar{x}(t) \sim L/\left[\mu'_0  \ln(t/\tau_{0}) \right]$ as $t\rightarrow \infty $.
It is clear that particles move ultra-slowly towards $x=0$ since the mean position of particles decreases to zero logarithmically.

\textit{Monte Carlo Simulations}. To verify the asymptotic density (\ref{longtsolfde}), we perform Monte Carlo simulations of the following random walk. There are $k$ boxes equally spaced between $x=0$ and $x=L$ with each box $i$ having length $a = \frac{L}{k}$. A particle resides in box $i$ for a random residence time $T$ drawn from a PDF, $\psi_{\mu_i}(\tau) = -\frac{\partial}{\partial \tau}E_{\mu_i}(-(\tau/\tau_0)^{\mu_i})$ (details in \cite{fulger2008monte}) where $\mu_i$ is a  discrete sampling of a linearly increasing function, $\mu(x) = \mu(0) + \left[\mu(L)-\mu(0)\right]x/L$ and $\tau_0$ is the time scale as before. After waiting for time $T$ it hops right with probability $r_i$ or left with probability $1-r_i$, except for  when the particle occupies state $i = 1$ or $k$. At the boundaries, the particles are reflected. The escape rate from the box $i$ is $I_i(t)=\tau_0^{-\mu_i}D^{1-\mu_i}_tp_i(t)$ \cite{henry2010fractional,fedotov2012subdiffusive}. The master equation can be written as
\begin{equation}
\raggedright
\begin{aligned}
	\frac{dp_i(t)}{dt} = & -\frac{1}{\tau_0^{\mu_i}}D^{1-\mu_i}_tp_i(t) + \\& \frac{1-r_{i+1}}{\tau_0^{\mu_{i+1}}}D^{1-\mu_{i+1}}_tp_{i+1}(t) + \frac{r_{i-1}}{\tau_0^{\mu_{i-1}}}D^{1-\mu_{i-1}}_tp_{i-1}(t)
	\end{aligned}
\end{equation}
where $p_i(t)$ is the probability that a particle occupies state $i$ at time $t$ \cite{henry2010fractional,fedotov2012subdiffusive}. In the continuous limit, this master equation for symmetric random walks, $r_i = 0.5$, reduces to the fractional diffusion equation (\ref{sdvofde}). For an asymmetric random walk, this master equation reduces to the fractional Fokker-Planck equation (\ref{voffpe}). 

Figure \ref{fig:fig1} shows the normalised histograms for $N=10^4$ particles performing the symmetric random walk with an uniform initial distribution; $r_i =0.5$, $L = 1$, $k=50$, $\tau_0 = 10^{-3}$ and $\mu_i = 0.4 + 0.5(i-1)/(k-1)$. One can see excellent agreement between the asymptotic solution (dashed line) and Monte Carlo simulations. The inset in Fig. \ref{fig:fig1} illustrates numerical confirmation of the ultra-slow logarithmic aggregation of particles at $x=0$ as predicted by (\ref{longtsolfde}). Furthermore, it shows the power-law decay of the PDF: $\ln(t)/t^{-\Delta \mu(x)}$ for $x\neq 0$.

To demonstrate numerically the dominance of the advection term involving the fractional operator (\ref{modifiedRLderivative}) over the standard advection in the variable-order fractional Fokker-Planck equation (\ref{voffpe}), we perform Monte Carlo simulations for an asymmetric random walk. We use $r_i = \frac{1}{2}+\frac{0.5}{k}\left( 0.5 - \frac{i}{k} \right)$
corresponding to the drift function $v(x) = 1-2x$ in Eq. (\ref{voffpe}) \cite{fedotov2012subdiffusive}. The motivation behind using this form of $r_i$ is to create advection that pushes particles to the center of the domain, $0<x<1$. For all other parameters, we use the same as in Fig.\ref{fig:fig1}. Figure \ref{fig:fig2} shows that at intermediate time, $t=5$, there is the formation of a Boltzmann-like distribution with the peak at the center of the domain. However, in the long-time limit, when $t=10^4$, the advection term involving the fractional operator (\ref{modifiedRLderivative}) is completely dominant and the asymptotic particle distribution corresponds to Eq. (\ref{longtsolfde}). If we approximate the non-uniform exponent by its mean value $\bar{\mu} = \frac{1}{L}\int_{0}^{L} \mu(x)dx$, then the asymptotic behavior of $p(x,t)$ will be very misleading because $p(x,t)$ approaches the Boltzmann distribution (see inset in Fig. \ref{fig:fig2}).

\textit{Summary}. We have obtained the asymptotic representation of the solution of the space-dependent variable-order fractional diffusion equation, which has remained unsolved since it was proposed in 2005 \cite{chechkin2005fractional}. We show that this solution remains valid for the fractional Fokker-Planck equation. It has been confirmed by direct numerical simulation of underlying anomalous CTRW. This asymptotic form describes the ultra-slow spatial aggregation of subdiffusive particles, which has no analogue in widely used classical advection-diffusion models. This new anomalous mechanism is generated by the space dependence of the fractional exponent, which leads to a new advection term involving a logarithmic modification of the Riemann-Liouville derivative. The unusual property of this advection is that it is always dominant over diffusion and standard drift regardless of the value of the gradient $\partial p /\partial x$ at long times. 

Experiments and analysis of empirical intracellular lysosome distribution \cite{ba2018whole} provides a possible basis for the formation of spatially non-uniform organelle distribution formation. The anomalous mechanism presented in this Letter is obviously not a complete theory to describe the non-uniform distribution of intracellular organelles. There are many other interactions and phenomena that occur in conjunction. Two primary additional phenomena that will affect this pattern is the superdiffusion generated by motor protein transport of organelles \cite{chen2015memoryless,fedotov2018memory,korabel2018non} and the non-linear interaction of subdiffusive organelles \cite{straka2015transport} such as the lysosome tethering to the endoplasmic reticulum observed in \cite{ba2018whole}. Furthermore, there are several other mechanisms, such as viscoelasticity and diffusion in labyrinthine environments, that lead to subdiffusive motion of organelles (see the excellent review \cite{sokolov2012models}). Including these additional effects in future works should provide a more physical and accurate model of organelle organization in the cell. 

\begin{acknowledgments}
	The authors acknowledge financial support from the EPSRC Grant No. EP/J019526/1 and the Wellcome Trust Grant No. 215189/Z/19/Z. The authors would like to thank V. J. Allan, M. Johnston, N. Korabel, H. Stage and T. Waigh for useful discussions.
\end{acknowledgments}

\bibliography{real}


\end{document}